# Modelo de Quarks e sistemas multiquarks

(Quark Model and multiquark system)

Cristiane Oldoni da Silva[1] e Paulo Laerte Natti[2]

*[1]Departamento de Física, Universidade Estadual de Londrina, Londrina, PR – Brasil*
*[2]Departamento de Matemática, Universidade Estadual de Londrina, Londrina, PR – Brasil*

A descoberta de muitas partículas, especialmente nos anos 50, com o surgimento de aceleradores, causou a busca por um modelo que descrevesse de forma simplificada o conjunto de partículas conhecidas. O Modelo de Quarks, baseado nas estruturas matemáticas da teoria de grupos, forneceu no início dos anos 60 uma descrição simplificada da matéria hadrônica conhecida, propondo que três partículas, chamadas quarks, dariam origem a todos os hádrons observados. Este modelo foi capaz de prever a existência de partículas posteriormente detectadas, confirmando sua consistência. Extensões do Modelo de Quarks foram implementadas no início dos anos 70 com o objetivo de descrever partículas observadas que eram estados excitados das partículas fundamentais e outras que apresentavam novos números quânticos (sabores). Recentemente, estados exóticos do tipo tetraquarks e pentaquarks, chamados também de sistemas multiquarks, já previstos pelo Modelo de Quarks, foram observados, o que renovou o interesse na forma como os quarks estão confinados dentro dos hádrons. Neste artigo apresentamos uma revisão do Modelo de Quarks e uma discussão sobre estes novos estados exóticos.
**Palavras-chave:** Modelo de Quarks, multiquarks, confinamento.

The discovery of many particles, especially in the 50's, when the firsts accelerators appeared, caused the searching for a model that would describe in a simple form the whole of known particles. The Quark Model, based in the mathematical structures of group theory, provided in the beginning of the 60's a simplified description of hadronic matter already known, proposing that three particles, called quarks, would originate all the observed hadrons. This model was able to preview the existence of particles that were later detected, confirming its consistency. Extensions of the Quark Model were made in the beginning of the 70's, focusing in describing observed particles that were excited states of the fundamental particles and others that presented new quantum numbers (flavors). Recently, exotic states as tetraquarks and pentaquarks types, also called multiquarks systems, previewed by the model, were observed, what renewed the interest in the way as quarks are confined inside the hadrons. In this article we present a review of the Quark Model and a discussion on the new exotic states.

---

[1] E-mail: crisoldoni@yahoo.com.br
[2] E-mail: plnatti@uel.br

# I Introdução

Desde a antiguidade, os filósofos já se preocupavam em encontrar os elementos básicos responsáveis pela constituição da matéria. Os filósofos gregos pré-socráticos imaginavam que os diversos materiais do mundo fossem formados pelos quatro elementos fogo, água, terra e ar, como pensava Empédocles (494 - 434 a.C.), ou mesmo que um único elemento básico, por exemplo, a água, pudesse se transformar em qualquer coisa, conforme imaginava Tales de Mileto (624 - 548 a.C.). Demócrito (460 - 370 a.C.) formulou a idéia de que todos os corpos são formados por pequenas unidades, fundamentais, às quais ele atribui o nome átomos.

Em 1808, John Dalton (1766 - 1844) propôs um modelo atômico, no qual as diversas substâncias eram combinações de átomos de hidrogênio, carbono, ferro etc..., explicando a essência de toda matéria encontrada na natureza.

Em 1869, Dimitri Mendeleev (1834 - 1907) classifica os elementos químicos conhecidos, de acordo com as suas propriedades, em uma tabela, indicando que os átomos eram compostos por "blocos de construção" ainda menores, e que esses blocos em diferentes combinações determinavam as propriedades dos elementos químicos. Devido à periodicidade das propriedades, ele foi capaz de prever a existência de elementos ainda desconhecidos, que vieram a ser descobertos algum tempo depois.

John Joseph Thomson (1856 - 1940), realizando experiências com tubos de raios catódicos, mostrou em 1897 que o feixe de raios catódicos era composto de partículas com carga elétrica negativa, determinando a velocidade e a razão carga-massa destes corpúsculos, posteriormente chamados elétrons. Em 1898 propôs um modelo, que ficou conhecido como "pudim de passas", onde os átomos não seriam mais indivisíveis, mas formados por uma "pasta" de carga positiva com cargas negativas (passas) homogeneamente distribuídas nela.

Em 1886, Eugene Goldstein (1850 – 1930), também realizando pesquisas com tubos de raios catódicos, mas com catodos perfurados, observou feixes luminosos no sentido oposto ao dos raios catódicos, os quais foram chamados de raios canais. Posteriormente, verificou-se que os raios canais eram compostos por partículas positivas, às quais foi dado o nome de próton em 1904. Desta forma o modelo de Thomson se tornou incompleto, pois não levava em consideração a existência dos prótons.

Ernest Rutherford (1871 - 1937), em 1911, utilizando partículas alfa, mostrou que o núcleo atômico deveria ser extremamente denso e formado por cargas positivas, prótons, enquanto os elétrons estariam distribuídos ao redor do núcleo, o que faria com que os átomos tivessem grandes espaços vazios.

Para que houvesse avanço na compreensão dos modelos atômicos, foram essenciais a descoberta e utilização dos raios-X, e das radiações alfa, beta e gama. Experiências realizadas por James Chadwick (1891 - 1974) em 1932, envolvendo partículas alfa, indicaram mais um componente da família atômica: o nêutron. Assim, todos os elementos químicos presentes na tabela periódica passaram a ser entendidos como combinações de somente três partículas diferentes: prótons, nêutrons e elétrons. Com o aperfeiçoamento da câmera de nuvens, criada por Charles Thomson Rees Wilson (1869 - 1959) em 1911, foram observadas novas partículas, o pósitron, por exemplo, em 1931, que é a antipartícula do elétron. Outras partículas foram sendo detectadas, dentre elas a partícula de Hideki Yukawa (1907 - 1981), chamada de píon, em 1947, através de estudos de chuveiros atmosféricos gerados por raios cósmicos. Entre 1950 e 1960 surgiram aceleradores de partículas que permitiram que reações muito mais energéticas fossem realizadas. Uma nova propriedade das partículas foi descoberta, a estranheza, que é uma característica intrínseca da partícula, como a sua massa e sua carga elétrica, por exemplo. Desta forma a lista de partículas conhecidas cresceu substancialmente. Com a descoberta de mais partículas, começou a ficar claro que existia uma nova ordem em meio à abundância de partículas observadas.

No final dos anos cinqüenta buscava-se uma simetria, como as periodicidades identificadas por Mendeleev, que ordenasse o conjunto de partículas conhecidas [1]. O modelo de Sakata (1956), baseado na estrutura matemática do grupo SU(3), ao estender a idéia de Fermi e Yang, considerando o próton, o nêutron e também a partícula estranha $\Lambda$ como as partículas fundamentais, foi um primeiro passo que permitiu uma descrição simplificada da matéria hadrônica. Em 1961, Gell-Mann e Ne'eman, independentemente, observaram que os 8 bárions conhecidos, o próton e o nêutron entre eles, com spin igual a $\frac{1}{2}$, apresentavam um padrão quando esquematizada a estranheza destes em função de suas cargas. Os mésons com spin 0 também apresentavam este padrão, ou seja, a formação de octetos de partículas. Esta classificação dos bárions e mésons em octetos ficou conhecida como Caminho Óctuplo ou Caminho do Octeto. Em 1964, Gell-Mann e Zweig, independentemente, propuseram um modelo onde três partículas, com número bariônico e carga fracionários, chamadas quarks, dariam origem a todos os hádrons observados, o chamado Modelo de Quarks, explicando inclusive o padrão dos octetos . A descoberta, ainda em 1964, da partícula $\Omega^-$ formada de três quarks estranhos, a qual em 1962 havia sido prevista por Gell-Mann, confirmou a consistência do Modelo de 3 Quarks. Atualmente já foram detectados seis tipos diferentes de quarks e para cada um deles sua respectiva antipartícula [2].

A partir dos anos 70, como previsto pelo Modelo de Quarks, começaram a ser observados estados hadrônicos excitados em momento angular, excitação radial e outras excitações, semelhantes às excitações anteriormente observadas nos núcleos atômicos, das partículas que preenchiam as configurações mais baixas previstas pelo Modelo de Quarks [3], o que elevou dramaticamente o número de hádrons observados. Enfim, a partir de 2002, estados hadrônicos exóticos formados por quatro quarks, chamados tetraquarks, e por cinco quarks, chamados pentaquarks, foram observados [4].

Neste artigo vamos revisitar o Modelo de Quarks e discutir os novos modelos para a força entre os quarks. Na seção II apresentamos, de forma breve, o modelo. Nas seções III, IV e V desenvolvemos uma descrição das propriedades das partículas através da simetria dos grupos SU(N). Na seção VI, o Modelo de Quarks é estendido para descrever excitações observadas dos estados fundamentais das partículas. Finalmente, na seção VII discutimos as recentes observações de partículas exóticas, também chamadas de sistemas multiquarks.

## II Modelo de Quarks

Os hádrons, palavra de origem grega que significa pesado, são todas as partículas formadas por quarks. A interação responsável por manter os quarks unidos no núcleo atômico é a interação forte, assim como a interação eletromagnética mantém prótons e elétrons unidos formando átomos. Note que na escala do núcleo atômico, a interação forte é mais intensa que a interação eletromagnética. Experimentalmente, verifica-se que os quarks não se manifestam em estados livres [5]. Acredita-se que os quarks tenham massas de repouso extremamente elevadas e se liguem para formar hádrons com energias de ligação igualmente elevadas, de modo que somente em situações (colisões) extremamente energéticas eles se manifestem como partículas livres em pequenas distâncias. Este fenômeno é conhecido como confinamento dos quarks e está relacionado a um número quântico, como a carga elétrica, chamado carga de cor [6,7]. Neste modelo, as cores nas quais os quarks poderiam existir são, por convenção, vermelho ($R$ – $red$), azul ($B$ – $blue$) e amarelo ($Y$ – $yellow$) e as respectivas anticores. No entanto, ainda de acordo com o modelo, somente estados, hádrons, sem cor podem ser observados. Para que isto ocorra, somente é possível formar partículas com três quarks ($qqq$), um de cada cor, denominadas bárions, ou partículas com um quark e um antiquark ($q\bar{q}$), com uma cor e sua anticor, denominadas mésons. Esta teoria para a interação dos quarks, chamada Cromodinâmica Quântica [6,7], ou simplesmente QCD, foi verificada experimentalmente a partir de 1968, no Acelerador Linear de Stanford, Califórnia, SLAC, quando foram realizadas medições do espalhamento de elétrons altamente energéticos de $20\,GeVs$ sobre prótons, para grande momentos transferidos. Os resultados, indiretamente, confirmaram a existência dos quarks, pois indicaram colisões com partículas pontuais de cargas $+(2/3)e$ e $-(1/3)e$, onde $e$ é a carga fundamental, mostrando que o próton não é fundamental, mas sim formado por três partículas, como previsto no modelo dos quarks. Posteriormente, medidas no SLAC em 1972, da seção de choque do espalhamento $e^+e^- \to q\bar{q}$, proporcional ao número de cores, indicaram resultados compatíveis com a existência de três cores, confirmando as predições da QCD.

Existem muitas analogias entre a Eletrodinâmica Quântica [6,8], ou simplesmente QED, e a Cromodinâmica Quântica. Na QED temos a carga elétrica (positiva/negativa), enquanto na QCD temos as três cargas de cor (cor/anticor). Na QED o fóton é responsável pela interação eletromagnética entre partículas carregadas, enquanto na QCD oito glúons são responsáveis pela interação forte entre os quarks carregados de cor. Na QCD cores iguais se repelem, enquanto cores diferentes se atraem. Enfim, a grande diferença entre estas teorias é que os oito glúons carregam cor, enquanto o fóton não tem carga elétrica, isto faz com que os glúons interajam fortemente entre si. Esta última característica da QCD gera uma Lei de Gauss para a interação forte muito diferente da Lei de Gauss elétrica, consequentemente o comportamento da força forte com a distância torna-se também muito diferente daquele apresentado pela força elétrica.

Os quarks, assim como os elétrons, possuem um spin intrínseco igual a $1/2$ e, por convenção, paridade positiva. Os bárions, estados de três quarks, podem ter spin total ($S = S_1 + S_2 + S_3$) igual a $S = 1/2$ ou $S = 3/2$, ou seja, spins semi-inteiros, sendo, portanto, classificados como férmions (partículas que obedecem à estatística de Fermi-Dirac e ao Princípio de Exclusão de Pauli). Por outro lado, os mésons, estados de dois quarks, podem ter spin total ($S = S_1 + S_2$) igual a $S = 0$ ou $S = 1$, sendo classificados como bósons (partículas que obedecem à estatística de Bose-Einstein) [9,10]. Os quarks são férmions e, portanto, quarks idênticos não podem se combinar para formar bárions (segundo o princípio de exclusão), a menos que exista outro número quântico que os diferencie. Esse número é justamente a carga de cor.

Até o momento já foram observados seis tipos de quarks, também chamados sabores, que são o quark $u$ (*up*), o quark $d$ (*down*), o quark $s$ (*strange*), o quark $c$ (*charm*), o quark $b$ (*bottom*) e o quark $t$ (*top*) [5]. Por simetrias, como o número de léptons também é seis, ou seja, elétron e neutrino eletrônico, múon e neutrino muônico, tau e neutrino tauônico [6], e porque tanto os quarks como os léptons agrupam-se em famílias de dois elementos, acredita-se que o número de sabores (tipos) de quarks existentes deva ser par e não há evidências para que existam outros sabores de quarks [5,6].

Os quarks constituintes dos prótons e nêutrons são, respectivamente, $u+u+d$ e $u+d+d$. Como existe diferença de carga elétrica entre prótons e nêutrons, admite-se que o quark $u$ tenha carga fracionária $+\frac{2}{3}e$, enquanto o quark $d$ tenha carga $-\frac{1}{3}e$, onde $e$ é a carga fundamental. Desta forma, a carga do próton é $+1$ e do nêutron é $0$. Por outro lado, os quarks constituintes dos mésons $\pi^+$ e $\pi^-$ são $u+\bar{d}$ e $\bar{u}+d$, respectivamente. Observe que os mésons $\pi^+$ e $\pi^-$ têm seus quarks e antiquarks trocados, de modo que o méson $\pi^-$ é antipartícula do méson $\pi^+$ e vice-versa. Na próxima seção, considerando apenas dois sabores de quarks, ou seja, os quarks $u$ e $d$, identificaremos as partículas fundamentais, bárions e mésons, que são estados ligados destes quarks.

## III SU(2): Partículas com dois sabores de quarks

O conceito de simetria é fundamental em Física Teórica e a linguagem matemática natural para descrever simetrias (periodicidades) é a Teoria de Grupos [11], que foi utilizada no desenvolvimento do modelo de quarks, em especial o SU(N), responsável por gerar os padrões observados nas tabelas de partículas.

Prótons e nêutrons são partículas muito parecidas, com massas muito próximas, diferindo por suas cargas elétricas, de modo que elas foram interpretadas como manifestações de uma mesma partícula, chamada núcleon, situação análoga aos dois estados de um elétron (spin para baixo e spin para cima). A interação eletromagnética seria responsável pela quebra de simetria do núcleon, e, portanto, pela diferença de massa entre prótons e nêutrons. A estrutura matemática usada para discutir a similaridade entre nêutrons e prótons, chamada isospin, é uma cópia do formalismo utilizado para descrever os estados de spin do elétron, ou seja, o grupo SU(2). Os geradores de SU(2), ou seja, as matrizes que geram as transformações do grupo SU(2), são as matrizes de Pauli [10-12],

$$\sigma_1 = \begin{pmatrix} 0 & 1 \\ 1 & 0 \end{pmatrix} \quad \sigma_2 = \begin{pmatrix} 0 & -i \\ i & 0 \end{pmatrix} \quad \sigma_3 = \begin{pmatrix} 1 & 0 \\ 0 & -1 \end{pmatrix}, \quad (1)$$

que formam uma base, de modo que qualquer matriz unitária, $2 \times 2$, com determinante $+1$, pode ser escrita como uma combinação linear de $\sigma_1$, $\sigma_2$, $\sigma_3$ e da matriz identidade.

Consideremos a interação de dois núcleons (prótons e/ou nêutrons) de isospin $I = \frac{1}{2}$. Como as projeções são $I_3 = \pm \frac{1}{2}$, temos estados núcleon-núcleon [6,12] com isospin total $I = 1$ ou $I = 0$, ou seja,

Tripleto de estados com $I = 1$
$$|I=1, I_3=1\rangle = |\tfrac{1}{2}, \tfrac{1}{2}\rangle \otimes |\tfrac{1}{2}, \tfrac{1}{2}\rangle = |p\rangle|p\rangle$$
$$|I=1, I_3=0\rangle = \sqrt{\tfrac{1}{2}}(|p\rangle|n\rangle + |n\rangle|p\rangle) \quad (2)$$
$$|I=1, I_3=-1\rangle = |\tfrac{1}{2}, -\tfrac{1}{2}\rangle \otimes |\tfrac{1}{2}, -\tfrac{1}{2}\rangle = |n\rangle|n\rangle$$

Singleto de estados com $I = 0$
$$|I=0, I_3=0\rangle = \sqrt{\tfrac{1}{2}}(|p\rangle|n\rangle - |n\rangle|p\rangle). \quad (3)$$

Na realidade, a simetria entre prótons e nêutrons é devida à semelhança de massas dos quarks $u$ e $d$, $m_u \approx m_d \approx 350$ MeVs. Portanto, a simetria de isospin é uma simetria entre os quarks $u$ e $d$, e não entre os núcleons.

Por convenção, o quark $u = |\tfrac{1}{2}, +\tfrac{1}{2}\rangle$ é representado por um estado de isospin $I = \tfrac{1}{2}$ e projeção $I_3 = +\tfrac{1}{2}$, enquanto o quark $d = |\tfrac{1}{2}, -\tfrac{1}{2}\rangle$ é representado por um estado de isospin $I = \tfrac{1}{2}$ e projeção $I_3 = -\tfrac{1}{2}$. Da mesma forma como foi exposto anteriormente para núcleons, utilizando a notação $|I, I_3\rangle$, onde $I$ é o isospin total e $I_3$ é a projeção do isospin total, analisemos a interação entre quarks $u$ e $d$, formando estados com isospin total $I = 1$ ou $I = 0$,

Tripleto de estados com $I = 1$
$$|1,1\rangle = uu$$
$$|1,0\rangle = \sqrt{\tfrac{1}{2}}(ud + du) \quad (4)$$
$$|1,-1\rangle = dd$$

Singleto de estados com $I = 0$
$$|0,0\rangle = \sqrt{\tfrac{1}{2}}(ud - du), \quad (5)$$

onde a seguinte notação foi utilizada $|1,1\rangle = |½,½\rangle \otimes |½,½\rangle = |u\rangle \otimes |u\rangle = |u\rangle|u\rangle = uu$. Podemos representar simbolicamente esta interação por

$$\tfrac{1}{2} \otimes \tfrac{1}{2} = 1_S \oplus 0_A, \quad (6)$$

onde os índices $S$ e $A$ denotam estados simétricos e anti-simétricos por troca de partículas, respectivamente. Alternativamente, através da notação ($2S+1$), temos

$$2 \otimes 2 = 3 \oplus 1, \quad (7)$$

ou seja, dois tipos de quarks combinados com dois tipos de quarks geram um tripleto de estados simétricos e um estado singleto anti-simétrico [13]. Note que os estados formados por dois quarks, em (4) e (5), são coloridos, pois, de acordo com a QCD, somente estados formados por três quarks ($qqq$), um de cada cor, denominados bárions, ou estados formados por um quark e um antiquark ($q\bar{q}$), com uma cor e sua anticor, denominados mésons, são estados sem cor. Ainda, de acordo com a QCD, estados coloridos não são estados observáveis, pois não há partículas físicas que possam ser associadas a estes estados com carga e número bariônico semi-inteiros.

Para formar partículas observadas com três quarks, um de cada cor, como os bárions, é necessário combinar os estados de dois quarks dados em (4) e (5) com um terceiro quark. Como tínhamos quatro estados e vamos combinar cada um deles com um quark $u$ ou um quark $d$, obteremos oito possíveis estados. A construção detalhada destes estados pode ser encontrada nas referências [6,13,14]. Convidamos o leitor a construir estes 8 estados quânticos a partir dos estados quânticos de 2 quarks dados em (4) e (5). Para dois quarks tínhamos o isospin total $I=1$ ou $I=0$. Adicionando uma terceira partícula para os estados simétricos ($I=1$), teremos estados com isospin total $I=\tfrac{3}{2}$ ou $I=\tfrac{1}{2}$. Para o estado anti-simétrico ($I=0$) a adição de uma terceira partícula fornecerá apenas estados $I=\tfrac{1}{2}$. Os oito estados estão listados na Tabela 1 por suas projeções de isospin, onde utilizamos a notação $qqq = |q\rangle \otimes |q\rangle \otimes |q\rangle$, por simplicidade.

Tabela 1 – Estados ligados de três quarks para dois sabores de quarks [6].

| $I_3 = \tfrac{3}{2}$ | $I_3 = \tfrac{1}{2}$ | $I_3 = -\tfrac{1}{2}$ | $I_3 = -\tfrac{3}{2}$ |
|---|---|---|---|
| $uuu = \Delta^{++}$ | $\tfrac{1}{\sqrt{3}}(uud + udu + duu) = \Delta^+$ | $\tfrac{1}{\sqrt{3}}(ddu + dud + udd) = \Delta^0$ | $ddd = \Delta^-$ |
|  | $\tfrac{1}{\sqrt{6}}[(ud+du)u - 2uud] = p_S$ | $-\tfrac{1}{\sqrt{6}}[(ud+du)d - 2ddu] = n_S$ |  |
|  | $\tfrac{1}{\sqrt{2}}(ud-du)u = p_A$ | $\tfrac{1}{\sqrt{2}}(ud-du)d = n_A$ |  |

Escrevendo na forma simbólica, como fizemos para duas partículas [13]

$$\left(\tfrac{1}{2} \otimes \tfrac{1}{2}\right) \otimes \tfrac{1}{2} = \left(1 \otimes \tfrac{1}{2}\right) \oplus \left(0 \otimes \tfrac{1}{2}\right) = \left(\tfrac{3}{2}_S \oplus \tfrac{1}{2}_{SM}\right) \oplus \tfrac{1}{2}_{AM}$$

ou

$$(2 \otimes 2) \otimes 2 = (3 \otimes 2) \oplus (1 \otimes 2) = (4 \oplus 2) \oplus 2, \quad (8)$$

identificamos quatro estados simétricos ($S$) de isospin $\tfrac{3}{2}$ (primeira linha da Tabela 1), dois estados simétricos misturados ($SM$, simétricos com relação à troca dos dois primeiros quarks) de isospin $\tfrac{1}{2}$ (segunda linha da Tabela 1) e dois estados anti-simétricos misturados ($AM$) de isospin $\tfrac{1}{2}$ (terceira linha da Tabela 1).

A partícula $\Delta^{++}$, lê-se delta mais mais, da Tabela 1 é chamada de ressonância $\Delta^{++}$ [14]. Ela foi a primeira ressonância observada, por Fermi em 1952, ao medir um pico na seção de choque do espalhamento píon-próton com tempo de meia-vida de $\tau \approx 10^{-23} s$. Meias-vidas desta magnitude são características de reações envolvendo a interação forte. Ressonâncias são estados gerados, via interação forte, com spin, paridade, carga, massa, etc., bem definidos e que são entendidos como partículas elementares. Em termos dos constituintes quarkiônicos, a reação que gera a ressonância $\Delta^{++}$ pode ser representada por

$$p + \pi^+ \to \Delta^{++} + \pi^0 \to p + \pi^+$$
$$uud + \bar{d}u \to uuu + \bar{u}u \to uud + \bar{d}u.$$

Note que, neste espalhamento, um par de quarks $d\bar{d}$ é aniquilado e um par $u\bar{u}$ é criado num primeiro momento para gerar a ressonância $\Delta^{++}$ e, em seguida, temos um processo inverso que restaura as partículas iniciais. Outros processos podem ocorrer, como, por exemplo:

$$p + \pi^+ \to \Sigma^+ + K^+ \to p + \pi^+$$
$$uud + \bar{d}u \to uus + \bar{s}u \to uud + \bar{d}u.$$

Nesta última interação, a ressonância $\Sigma^+$, lê-se sigma mais, e o méson $K^+$, lê-se káon mais, são gerados devido à criação de um par de quarks estranhos $s\bar{s}$.

Enfim, as partículas (estados) designadas por $p_S$, $p_A$, $n_S$ e $n_A$, na Tabela 1, são os estados fundamentais e excitados de prótons e nêutrons, respectivamente, por troca de quarks. Em analogia com estados descritos por suas projeções de spin, estes estados seriam estados emparelhados e desemparelhados, respectivamente. Enfim, para que as funções de onda totais destas partículas fermiônicas sejam antissimétricas, devemos lembrar que os quarks possuem também o número quântico de cor, que assume os três possíveis valores $R$, $B$, e $Y$, de modo que a função de onda de cor pode ser simétrica ou antissimétrica por troca destes números quânticos. Por exemplo, para bárions (três quarks), uma função de onda de cor antissimétrica [6] é dada por

$$(qqq)_{\text{Singleto de Cor}} = \sqrt{\tfrac{1}{6}}(RYB - RBY + BRY - BYR + YBR - YRB).$$

Na próxima seção iremos estender este formalismo ao estudo de estados ligados de quarks, bárions e mésons, quando são considerados três possíveis sabores para os mesmos.

## IV SU(3): Partículas com três sabores de quarks

Com a descoberta de novas partículas no final dos anos cinquenta, percebeu-se que algumas partículas muito massivas, que deveriam decair rapidamente, não decaíam. Essa manifestação de estabilidade foi interpretada, naquele período, como devida a um novo tipo de quark, chamado quark $s$, com um novo número quântico que devia ser conservado nas reações, chamado estranheza [1,14] e denotado por $E$. Normalmente a estranheza de uma partícula é denotada por $S$, o que pode ser confundido com o seu estado de spin, também denotado por $S$, e, portanto, neste texto, denotaremos o estado de estranheza de uma dada partícula por $E$. O novo número quântico também pode ser representado pela hipercarga [1,14], $Y = B + E$, conforme a Figura 1, onde $B$ é o número bariônico. Como são necessários três quarks para formar um bárion, cada quark tem número bariônico $\tfrac{1}{3}$, enquanto antiquarks têm número bariônico $-\tfrac{1}{3}$. Note também que a carga dos quarks $u$, $d$ e $s$, dada por $Qe$, onde $e$ é a carga fundamental e $Q = I_3 + Y/2$, pode ser dada em termos da hipercarga e da projeção de isospin do quark considerado.

Os números quânticos dos quarks $u$, $d$ e $s$ são dados na Tabela 2. Observe que os estados dos quarks citados na Figura 1 e na Tabela 2 podem ser caracterizados por dois números quânticos, a projeção de isospin $I_3$ e a estranheza $E$. A carga e a hipercarga são dadas em termos destes números quânticos. Neste sentido, a carga elétrica de um quark é uma manifestação de seus estados de isospin e estranheza, lembrando que todos os quarks têm número bariônico $B = \tfrac{1}{3}$.

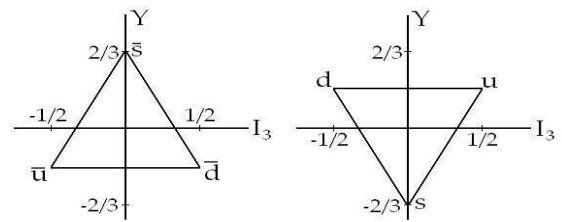

Figura 1 – SU(3): tripletos de quarks e antiquarks com os respectivos números quânticos de hipercarga $Y$ e projeção de isospin $I_3$.

Tabela 2 – Números quânticos dos quarks $u$, $d$ e $s$.

| Quark | Spin | $B$ | $Q$ | $I_3$ | $E$ | $Y$ |
|---|---|---|---|---|---|---|
| $u$ | $\tfrac{1}{2}$ | $\tfrac{1}{3}$ | $+\tfrac{2}{3}$ | $+\tfrac{1}{2}$ | 0 | $+\tfrac{1}{3}$ |
| $d$ | $\tfrac{1}{2}$ | $\tfrac{1}{3}$ | $-\tfrac{1}{3}$ | $-\tfrac{1}{2}$ | 0 | $+\tfrac{1}{3}$ |
| $s$ | $\tfrac{1}{2}$ | $\tfrac{1}{3}$ | $-\tfrac{1}{3}$ | 0 | $-1$ | $-\tfrac{2}{3}$ |

Com a existência de um número quântico aditivo, devido a um novo tipo de quark, em adição a $I_3$, seria natural tentar acomodar a nova simetria em um grupo maior. A partir de uma generalização da simetria de isospin, verificou-se que o formalismo utilizado para descrever os estados ligados (interações) possíveis destas partículas deveria ser unitário e descrito pelo grupo SU(3). Os geradores deste grupo são as oito matrizes de Gell-Mann [11,13]

$$\lambda_1 = \begin{pmatrix} 0 & 1 & 0 \\ 1 & 0 & 0 \\ 0 & 0 & 0 \end{pmatrix} \quad \lambda_2 = \begin{pmatrix} 0 & -i & 0 \\ i & 0 & 0 \\ 0 & 0 & 0 \end{pmatrix} \quad \lambda_3 = \begin{pmatrix} 1 & 0 & 0 \\ 0 & -1 & 0 \\ 0 & 0 & 0 \end{pmatrix}$$

$$\lambda_4 = \begin{pmatrix} 0 & 0 & 1 \\ 0 & 0 & 0 \\ 1 & 0 & 0 \end{pmatrix} \quad \lambda_5 = \begin{pmatrix} 0 & 0 & -i \\ 0 & 0 & 0 \\ i & 0 & 0 \end{pmatrix} \quad \lambda_6 = \begin{pmatrix} 0 & 0 & 0 \\ 0 & 0 & 1 \\ 0 & 1 & 0 \end{pmatrix}$$

$$\lambda_7 = \begin{pmatrix} 0 & 0 & 0 \\ 0 & 0 & -i \\ 0 & i & 0 \end{pmatrix} \quad \lambda_8 = \frac{1}{\sqrt{3}} \begin{pmatrix} 1 & 0 & 0 \\ 0 & 1 & 0 \\ 0 & 0 & -2 \end{pmatrix}. \quad (9)$$

Com três sabores de quarks ($u$, $d$ e $s$) para formar estados de dois quarks, teremos nove combinações possíveis, que podem ser separadas em seis estados simétricos e três anti-simétricos [6,13]. Estes estados são dados na Tabela 3. Assim, podemos escrever simbolicamente

$$3 \otimes 3 = 6 \oplus 3. \quad (10)$$

Lembramos que tais estados de dois quarks não são observados fisicamente, pois de acordo com a QCD somente estados de três quarks (bárions) e estados quark-antiquark (mésons) são observáveis.

Tabela 3 – Estados ligados (não-físicos) de dois quarks para três sabores de quarks.

| Simétricos | Anti-simétricos |
|---|---|
| $uu$ | |
| $\frac{1}{\sqrt{2}}(ud + du)$ | $\frac{1}{\sqrt{2}}(ud - du)$ |
| $dd$ | |
| $\frac{1}{\sqrt{2}}(us + su)$ | $\frac{1}{\sqrt{2}}(us - su)$ |
| $ss$ | |
| $\frac{1}{\sqrt{2}}(ds + sd)$ | $\frac{1}{\sqrt{2}}(ds - sd)$ |

Identifiquemos os mésons fundamentais que são estados ligados do tripleto de quarks ($u$, $d$ e $s$). Mésons são estados ligados de quark-antiquark com número bariônico $B = 0$. Os possíveis spins totais de um estado $q\bar{q}$ são $S = 0$ (quarks antiparalelos) ou $S = 1$ (quarks paralelos). Se o momento angular deste estado é $L$, então o momento angular total $J$ do méson assume os possíveis valores $|L-S| < J < |L+S|$. Define-se para os estados mesônicos a paridade ou paridade-P e a conjugação de carga ou paridade-C, respectivamente, como

$$P = (-1)^{L+1} \quad (11)$$
$$C = (-1)^{L+S}. \quad (12)$$

A operação de paridade é uma reflexão de todas as coordenadas através da origem, de modo que as partículas cuja função de onda muda de sinal sob a transformação $P$ são ditas terem paridade intrínseca ímpar, enquanto as que não mudam, paridade par. A operação de conjugação de carga corresponde à troca de uma partícula por sua antipartícula.

Vamos utilizar os operadores acima definidos para classificar os mésons em multipletos $J^{PC}$. Os estados de $L = 0$ são os multipletos pseudo-escalares ($J^{PC} = 0^{-+}$), quando $S = 0$, e os multipletos vetoriais ($J^{PC} = 1^{--}$), quando $S = 1$. Para os três sabores $u$, $d$ e $s$, temos também, obviamente, nove estados ligados quark-antiquark, no entanto agora agrupados em um octeto e num singleto de estados mesônicos. Note que para estados ligados quark-quark, com três sabores, tínhamos um sexteto e um tripleto de estados, veja a equação (10). A construção destes estados quânticos, que formam uma base neste espaço de estados, é efetuada nas referências [6,11,13]. Simbolicamente,

$$3 \otimes \bar{3} = 8 \oplus 1. \quad (13)$$

Os nonetos mesônicos são mostrados na Figura 2. Na Figura 2a apresentamos os estados experimentais quark-antiquark com momento angular total $J = 0$, que formam o multipleto $J^{PC} = 0^{-+}$ ($L = 0$, $S = 0$), e em 2b apresentamos os estados experimentais quark-antiquark com momento angular total $J = 1$, que formam o multipleto $J^{PC} = 1^{--}$ ($L = 0$, $S = 1$).

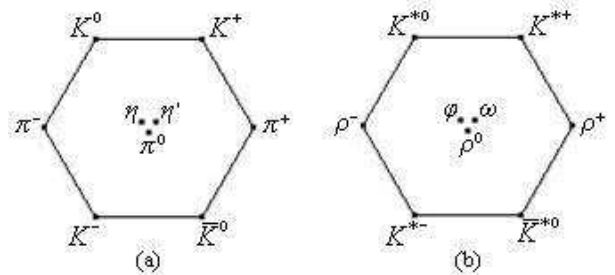

Figura 2 – Nonetos mesônicos: (a) Noneto mesônico pseudo-escalar $J^{PC} = 0^{-+}$. (b) Noneto mesônico vetorial $J^{PC} = 1^{--}$.

A representação destes estados, em termos dos três sabores de quarks considerados, é dada na Tabela 4. As partículas do multipleto $1^{--}$ podem ser entendidas como estados excitados de emparelhamento de spins das partículas do multipleto $0^{-+}$.

Consideremos agora as representações dos bárions, estados $qqq$, em termos de três sabores de quarks. Neste caso temos 27 estados que podem ser agrupados em quatro multipletos. Simbolicamente,

$$3 \otimes (3 \otimes 3) \rightarrow 3 \otimes (6 \oplus 3)$$
$$\rightarrow (10_S \oplus 8_{SM}) \oplus (8_{AM} \oplus 1). \quad (14)$$

Tabela 4 – Estados ligados mesônicos $q\bar{q}$ para três sabores de quarks.

| Partículas | Estados |
|---|---|
| $\pi^0$, $\rho^0$ | $(u\bar{u} - d\bar{d})/\sqrt{2}$ |
| $\pi^+$, $\rho^+$ | $u\bar{d}$ |
| $\pi^-$, $\rho^-$ | $-d\bar{u}$ |
| $\eta$ | $(u\bar{u} + d\bar{d} - 2s\bar{s})/\sqrt{6}$ |
| $\eta'$ | $(u\bar{u} + d\bar{d} + s\bar{s})/\sqrt{3}$ |
| $\omega$ | $(u\bar{u} + d\bar{d})/\sqrt{2}$ |
| $\varphi$ | $s\bar{s}$ |
| $K^+$, $K^{*+}$ | $u\bar{s}$ |
| $K^0$, $K^{*0}$ | $d\bar{s}$ |
| $\overline{K}^0$, $\overline{K}^{*0}$ | $-s\bar{d}$ |
| $K^-$, $K^{*-}$ | $s\bar{u}$ |

A representação dos multipletos dados em (14) é mostrada na Figura 3.

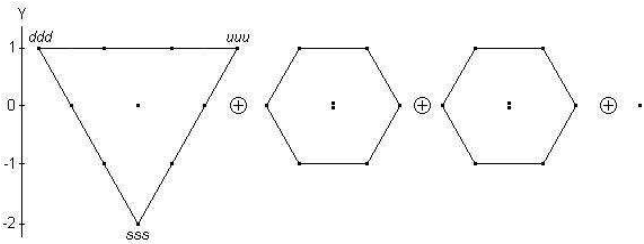

Figura 3 – Os multipletos $qqq$ de SU(3).

Quando $L = 0$, a Figura 4 identifica cada estado da Figura 3 com a respectiva partícula observada experimentalmente. Utilizamos a notação $J^P$ para caracterizar os multipletos deste caso, sendo $J$ o momento angular total e $P$ a paridade das partículas. Note que o octeto $J^P = \tfrac{3}{2}^-$ ausente na Figura 4 é formado por partículas que possuem os mesmos estados do octeto $J^P = \tfrac{1}{2}^+$ mostrado, e que podem ser entendidas como estados excitados de emparelhamento de spins das partículas do multipleto $J^P = \tfrac{1}{2}^+$.

A representação dos estados do decupleto, em termos dos três sabores $u$, $d$ e $s$ dos quarks, quando $L = 0$, é descrita na Tabela 5. A representação dos estados dos octetos são dados na Tabela 6 e o estado singleto é dado, explicitamente, por

$$\Lambda = \frac{1}{\sqrt{6}}\left[s(du - ud) + (usd - dsu) + (du - ud)s\right].$$

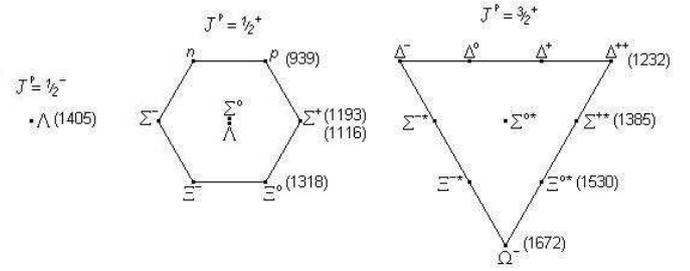

Figura 4 – Singleto $\tfrac{1}{2}^-$, octeto $\tfrac{1}{2}^+$ e decupleto $\tfrac{3}{2}^+$ de bárions com três sabores de quarks. Os valores entre parênteses correspondem às massas destas partículas em MeVs. O octeto $J^P = \tfrac{3}{2}^-$ ausente é formado pelos estados excitados de emparelhamento de spins do octeto $\tfrac{1}{2}^+$.

Tabela 5 – Representações dos estados simétricos $qqq$ pertencentes ao decupleto $J^P = \tfrac{3}{2}^+$, quando $L = 0$, para três sabores de quarks.

| Partícula | Estado simétrico |
|---|---|
| $\Delta^-$ | $ddd$ |
| $\Delta^0$ | $\dfrac{1}{\sqrt{3}}(ddu + dud + udd)$ |
| $\Delta^+$ | $\dfrac{1}{\sqrt{3}}(uud + udu + duu)$ |
| $\Delta^{++}$ | $uuu$ |
| $\Sigma^-*$ | $\dfrac{1}{\sqrt{3}}(dds + dsd + sdd)$ |
| $\Sigma^0*$ | $\dfrac{1}{\sqrt{6}}(dsu + uds + sud + sdu + dus + usd)$ |
| $\Sigma^+*$ | $\dfrac{1}{\sqrt{3}}(uus + usu + suu)$ |
| $\Xi^-*$ | $\dfrac{1}{\sqrt{3}}(dss + ssd + sds)$ |
| $\Xi^0*$ | $\dfrac{1}{\sqrt{3}}(uss + ssu + sus)$ |
| $\Omega^-$ | $sss$ |

Continuando nossa construção das partículas observadas em termos dos constituintes quarkiônicos, dados experimentais atualizados confirmam que já foram observadas partículas com cinco diferentes sabores de quarks. Em 1974, mésons que carregam o número quântico $c$, chamado *charm*, foram observados nos aceleradores de Brookhaven, em Long Island e no SLAC, na Califórnia, ambos nos Estados Unidos. Citamos algumas famílias de mésons com *charm*: mésons $D = c\bar{u}, \bar{c}u, c\bar{d}, \bar{c}d$ ($\approx 2450$ MeVs), mésons $D_s = c\bar{s}, \bar{c}s$ ($\approx 2500$ MeVs) e mésons $J/\psi = c\bar{c}$ (3100 MeVs). A partir destas medidas, e de modelos teóricos, tem-se que a massa do quark $c$ é da ordem de $m_c \approx 1870$ MeVs. Em 1977, mésons que carregam o número quântico $b$, chamado *bottom* ou *beauty*, foram observados no Fermilab, Chicago. Novamente citamos algumas famílias de mésons com *beauty*: mésons $B = b\bar{u}, \bar{b}u, b\bar{d}, \bar{b}d$ ($\approx 5270$ MeVs) e mésons $Y = b\bar{b}$ ($\approx 9460$ MeVs). Dados indicam que a massa do quark $b$ é da ordem de $m_c \approx 5280$ MeVs. Enfim, em 1996, no Fermilab, estimou-se que a massa do quark $t$ é da ordem de $m_t \approx 175000$ MeVs ou $m_t \approx 0,175$ TeVs. Salientamos que até o momento nenhum bárion ou méson, contendo o quark $t$, foi observado diretamente, devido à pequena meia-vida destas partículas. Uma tabela completa e atualizada destas famílias de estados mesônicos pesados pode ser encontrada em [5], no tópico *Quark Model*.

Consideremos o quarteto de sabores de quarks $u$, $d$, $s$ e $c$, como mostrado na Figura 5. Neste caso devemos estender a simetria de sabores para SU(4). Procedendo de forma análoga, obtemos para estados de três partículas interagentes (quarks), com quatro sabores possíveis, 64 estados de bárions agrupados em quatro multipletos. Simbolicamente,

$$4 \otimes 4 \otimes 4 \rightarrow 20_S \oplus 20_{SM} \oplus 20_{AM} \oplus 4_A. \quad (15)$$

A construção de todos estes 64 estados, e a identificação dos respectivos multipletos, torna-se extremamente trabalhosa, sendo, neste caso, melhor utilizarmos técnicas elegantes da teoria de grupos. Este estudo será descrito na próxima seção.

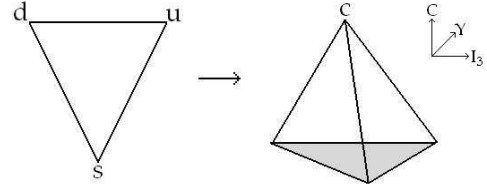

Figura 5 – Quarteto SU(4): o tripleto de quarks forma um quarteto com a adição do quark $c$.

Tabela 6 – Representações de simetrias misturadas para os estados $qqq$ pertencentes aos octetos $J^P = \frac{1}{2}^+$ e $J^P = \frac{3}{2}^-$, quando $L = 0$, para três sabores de quarks.

| Partícula | Estado misturado simétrico | Estado misturado anti-simétrico |
|---|---|---|
| $p$ | $\frac{1}{\sqrt{6}}[(ud+du)u - 2uud]$ | $\frac{1}{\sqrt{2}}(ud-du)u$ |
| $n$ | $-\frac{1}{\sqrt{6}}[(ud+du)d - 2ddu]$ | $\frac{1}{\sqrt{2}}(ud-du)d$ |
| $\Sigma^+$ | $\frac{1}{\sqrt{6}}[(us+su)u - 2uus]$ | $\frac{1}{\sqrt{2}}(us-su)u$ |
| $\Sigma^0$ | $\frac{1}{\sqrt{6}}\left[\frac{s(du+ud)}{\sqrt{2}} + \frac{dsu+usd}{\sqrt{2}} - \frac{2(du+ud)s}{\sqrt{2}}\right]$ | $\frac{1}{\sqrt{2}}\left[\frac{dsu+usd}{\sqrt{2}} - \frac{s(ud+du)}{\sqrt{2}}\right]$ |
| $\Sigma^-$ | $\frac{1}{\sqrt{6}}[(ds+sd)d - 2dds]$ | $\frac{1}{\sqrt{2}}(ds-sd)d$ |
| $\Lambda^0$ | $\frac{1}{\sqrt{2}}\left[\frac{dsu-usd}{\sqrt{2}} + \frac{s(ud-du)}{\sqrt{2}}\right]$ | $\frac{1}{\sqrt{6}}\left[\frac{s(du-ud)}{\sqrt{2}} + \frac{usd-dsu}{\sqrt{2}} - \frac{2(du-ud)s}{\sqrt{2}}\right]$ |
| $\Xi^-$ | $-\frac{1}{\sqrt{6}}[(ds+sd)s - 2ssd]$ | $\frac{1}{\sqrt{2}}(ds-sd)s$ |
| $\Xi^0$ | $-\frac{1}{\sqrt{6}}[(us+su)s - 2ssu]$ | $\frac{1}{\sqrt{2}}(us-su)s$ |

## V SU(N) e as Tabelas de Young

A obtenção dos estados ligados de quarks, $qqq$ e $q\bar{q}$, com $N > 3$, onde $N$ é o número de sabores, torna-se demasiadamente trabalhosa. Portanto seria proveitoso utilizar uma regra geral, a partir da qual fosse possível obter essas informações. O método das Tabelas de Young [6,11] permite a obtenção das representações para SU($N$) de forma rápida e elegante. Sendo $N$ o número de sabores

de quarks que estamos considerando, definimos a seguinte notação:

Representa o estado $N$

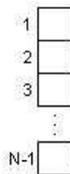

Representa o conjugado $N*$ ($N-1$ caixas empilhadas)

Exemplos:

N = 2 ▢    N* = 2 ▢
N = 3 ▢    N* = 3 ▯
N = 4 ▢    N* = 4 ▯

Tendo em vista essa representação simbólica, procedemos da seguinte forma para realizar a operação (interação) entre duas partículas, cada uma com $N$ sabores, através do método das Tabelas de Young [6,11]:

$$N \otimes N = \square \otimes \square = \square\square \oplus \begin{array}{c}\square\\\square\end{array} = A \oplus B \quad (16)$$

Portanto, para estados ligados de dois quarks, com $N$ sabores, representados pelo produto $N \otimes N$, temos as possíveis representações (multipletos) dadas pelos coeficientes racionais $A$ e $B$. Para determinar $A$ e $B$ devemos implementar dois procedimentos que identificam os respectivos numeradores e denominadores destes coeficientes.

– Procedimento para a determinação do numerador. Para um dado diagrama, representando o produto das representações de SU($N$), insere-se $N$ em cada caixa da diagonal principal, começando pelo canto superior esquerdo. Nas diagonais imediatamente acima e abaixo se insere $N+1$ e $N-1$, respectivamente. Nas próximas diagonais insere-se $N+2$ e $N-2$ e assim por diante, conforme mostrado no diagrama abaixo. O numerador do coeficiente do diagrama é o resultado da multiplicação de todas as quantidades dentro das caixas.

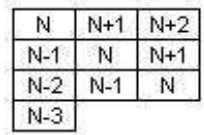

– Procedimento para a determinação do denominador. O número associado ao denominador é obtido através do Método dos Ganchos. O Método dos Ganchos [6,11,13] consiste em traçar linhas, através das caixas, que entram pela lateral direita do diagrama analisado. Quando entra na caixa, a linha gira 90º e desce por uma coluna até deixar o diagrama. O número total de caixas, pelas quais a linha passou, é o valor do gancho associado a estas caixas. Devem-se traçar todos os ganchos possíveis através das caixas do diagrama. O produto de todos os ganchos é o denominador.

Consideremos o caso já estudado na seção III de estados ligados de dois quarks com dois possíveis sabores. Nesta situação tínhamos obtido que $2 \otimes 2 = 3 \oplus 1$, onde $A = 3$ e $B = 1$. Calculemos estes coeficientes através do método das Tabelas de Young. Vejamos:

$$2 \otimes 2 = \square\square \oplus \begin{array}{c}\square\\\square\end{array}$$

sendo que

$$\square\square = \frac{\boxed{N\ N+1}}{2 \times 1 = 2} = \frac{N(N+1)}{2} = \frac{6}{2} = 3 \text{ pois } N = 2$$

e

$$\begin{array}{c}\square\\\square\end{array} = \frac{\boxed{\begin{array}{c}N\\N-1\end{array}}}{2 \times 1 = 2} = \frac{N(N-1)}{2} = \frac{2}{2} = 1.$$

Consideremos outro exemplo, já estudado na seção IV. Calculemos as representações de estados ligados de quark-antiquark quando $N = 3$. Neste caso temos que utilizar o conceito de diagrama conjugado $N* = \overline{N}$. Vejamos,

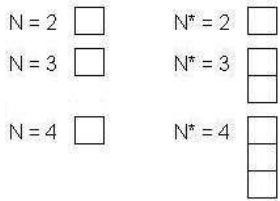

$$3 \otimes \bar{3} = \square \otimes \begin{array}{c}\square\\\square\end{array} = \begin{array}{cc}\square&\square\\\square&\end{array} \oplus \begin{array}{c}\square\\\square\\\square\end{array} = A \oplus B$$

Calculando os numeradores e denominadores, obtemos $A = 8$ e $B = 1$, ou seja, $3 \otimes \bar{3} = 8 \oplus 1$.

Enfim, podemos obter a representação irredutível de estados ligados de bárions (três quarks), para $N$ sabores,

combinando mais um estado com os dois estados que já haviam sido combinados em (16), ou seja,

$$(N \otimes N) \otimes N = \left( \square\square \oplus \begin{array}{c}\square\\\square\end{array} \right) \otimes \square = \square\square\square \oplus \begin{array}{c}\square\square\\\square\end{array} \oplus \begin{array}{c}\square\square\\\square\end{array} \oplus \begin{array}{c}\square\\\square\\\square\end{array}$$

Para obter o valor dos coeficientes devemos seguir os procedimentos já enunciados para calcular os numeradores e os denominadores de cada diagrama. Deixamos para o leitor esta tarefa. Utilizando esses procedimentos, o resultado geral para a interação de três quarks em SU($N$) é dado por

$$(N \otimes N) \otimes N = \left[ \frac{N(N+1)(N+2)}{6} \oplus \frac{(N-1)N(N+1)}{3} \right] \quad (17)$$

$$\oplus \left[ \frac{N(N-1)(N-2)}{6} \oplus \frac{(N-1)N(N+1)}{3} \right].$$

Através da equação (17) é possível obter os resultados já citados anteriormente para $(2 \otimes 2) \otimes 2$, $(3 \otimes 3) \otimes 3$ e $(4 \otimes 4) \otimes 4$, além de todas as representações irredutíveis para um $N$ qualquer. No modelo atual de quarks, considerando a existência de seis sabores de quarks, obtemos que a representação irredutível para estados bariônicos $qqq$ é dada em termos dos seguintes multipletos

$$(6 \otimes 6) \otimes 6 = (56 \oplus 70) \oplus (70 \otimes 20). \quad (18)$$

Devemos salientar que grande parte das partículas previstas em (18) ainda não foram observadas devido às limitações em energia, sensibilidade, etc., dos grandes acelerados atuais.

Na próxima seção o Modelo de Quarks será adaptado para descrever excitações observadas dos estados fundamentais das partículas descritas nas seções precedentes.

## VI Modelo de Quarks e Estados Excitados

Apesar de ser razoavelmente pequena a quantidade de bárions e mésons fundamentais formados pelos quarks mais leves ($u$, $d$ e $s$), a quantidade de partículas excitadas, constituídas por esses quarks, que são observadas, é imensa. Inicialmente, foram observadas partículas que preenchiam as configurações mais baixas previstas pelo modelo de quarks: estados quark-antiquark (mésons) ou estados de três quarks (bárions) com momento angular orbital $L = 0$. Estas são as partículas descritas na seção IV. Entretanto, começaram aparecer estados excitados destas partículas, fazendo com que o modelo fosse estendido para poder acomodar tais estados. Excitações de energias mais altas podem ser obtidas por excitação radial, excitação orbital e pela adição de pares quark-antiquark [13,15]. Estes estados excitados são instáveis ao decaimento forte e são observados experimentalmente como ressonâncias, mas são descritos como estados ligados no modelo de quarks.

Os estados excitados podem ser divididos em dois tipos, dependendo se a mudança no estado de cada par de quarks é em grau de liberdade interno ou espacial [13]. Os estados são caracterizados pelos números quânticos ($n$, $L$, $S$, $J$, $I$, $E$, etc.), que fornecem a excitação radial, o momento angular orbital, o spin, o momento angular total, o isospin, a estranheza, etc., destes. As excitações radiais e orbitais são expressas por uma coordenada relativa simples e uma variação de momento angular do quark ativo em relação ao resto do sistema, considerando que apenas um quark, em um bárion, por exemplo, é responsável pela transição que produz a ressonância sendo os outros dois apenas espectadores e permanecendo no mesmo estado inicial. Uma particularidade característica destas excitações é a alternância de paridade com aumento do momento angular orbital.

Para os mésons, a paridade-P e a paridade-C, já definidas em (11-12), a conjugação (troca) de carga-paridade e a paridade-G são operações definidas, respectivamente, por

$$CP = (-1)^{S+1} \quad (19)$$

$$G = (-1)^{L+S+I}, \quad (20)$$

onde o spin mesônico total $S$ pode ser $S = 0$ ou $S = 1$. Note que o isospin $I$ pode assumir os valores $I = 0$ ou $I = 1$, caso os constituintes quarkiônicos do méson considerado sejam apenas os quarks $u$ e $d$. No caso de mésons que contenham quarks $s$, $c$, $b$ e $t$ em sua constituição, o isospin $I$ também pode assumir o valor $I = \frac{1}{2}$, já que estes quarks têm isospin $I = 0$. A paridade-G é uma operação que combina a conjugação de carga com uma rotação de isospin. Tal quantidade é útil para mésons constituídos de quarks e seus próprios antiquarks, os quais apresentam carga de isospin nula ($I_z = 0$), e para os mésons com carga unitária de isospin ($I_z = 1$), como $u\bar{d}$ e $d\bar{u}$, pois é possível formular regras de seleção para estes sistemas isospin-carregados.

A Tabela 7 apresenta os estados mesônicos fundamentais e alguns excitados, em momento angular, formados pelos quarks $u$, $d$ e $s$. Nas duas primeiras linhas temos os mésons fundamentais já descritos na Tabela 4. Os primeiros estados excitados mesônicos, em momento angular, $L = 1$, são apresentados nas linhas 3 a 6 da Tabela 7, onde o sistema quark-antiquark apresenta um movimento relativo numa onda tipo-$p$. Estes mésons foram previstos teoricamente pelo Modelo de Quarks nos anos sessenta, inclusive suas massas, através da fórmula de Gell-Mann e Okubo [6,13], e, alguns deles, observados experimentalmente no início dos anos setenta [3,5]. Os próximos estados mesônicos excitados, em massa crescente, não presentes na Tabela 7, são aqueles onde o sistema quark-antiquarks está na primeira excitação radial ($n=2$) numa onda tipo-$s$ ($L=0$). Uma tabela atualizada dos estados mesônicos excitados pode ser encontrada em [5], no tópico *Quark Model*.

Tabela 7 – Mésons com $L=0$ e $L=1$. Entre parênteses são dadas as massas dos mésons em MeVs.

| $^{2S+1}L_J$ | $J^{PC}$ | $I=1$ | $I=\frac{1}{2}$ | $I=0$ | $I=0$ |
|---|---|---|---|---|---|
| $^3 0_1$ | $1^{--}$ | $\rho$ (770) | $K^*$ (890) | $\omega$ (782) | $\varphi$ (1020) |
| $^1 0_0$ | $0^{-+}$ | $\pi$ (140) | $K$ (495) | $\eta$ (550) | $\eta'$ (958) |
| $^3 1_2$ | $2^{++}$ | $a_2$ (1320) | $K_2^*$ (1430) | $f_2$ (1270) | $f_2'$ (1525) |
| $^3 1_1$ | $1^{++}$ | $a_1$ (1260) | $K_{1A}$ (1270) | $f_1$ (1285) | $f_1'$ (1420) |
| $^3 1_0$ | $0^{++}$ | $a_0$ (1450) | $K_0^*$ (1430) | $f_0$ (1370) | $f_0'$ (1710) |
| $^1 1_1$ | $1^{+-}$ | $b_1$ (1235) | $K_{1B}$ (1400) | $h_1$ (1170) | $E$ (1380) |

Tabela 8 – Alguns bárions, de diferentes multipletos, constituídos pelos quarks $u$, $d$ e $s$. Entre parênteses são dadas as massas dos mésons em MeVs.

| $J^P$ | $E=0$ | $E=-1$ | | $E=-2$ | $E=-3$ | Classificação $(SU(6), L^P)$ | SU(3) |
|---|---|---|---|---|---|---|---|
| | $I=\frac{1}{2}$ ou $\frac{3}{2}$ | $I=0$ | $I=1$ | $I=\frac{1}{2}$ | $I=0$ | | |
| $\frac{3}{2}^+$ | $\Delta$ (1232) | | $\Sigma$ (1385) | $\Xi$ (1530) | $\Omega$ (1672) | $(56, 0^+)$ | 10 |
| $\frac{1}{2}^+$ | $N$ (939) | $\Lambda$ (1116) | $\Sigma$ (1193) | $\Xi$ (1318) | | $(56, 0^+)$ | 8 |
| $\frac{3}{2}^-$ | $N$ (1515) | $\Lambda$ (1690) | $\Sigma$ (1660) | $\Xi$ (1850) | | $(70, 1^-)$ | 8 |
| $\frac{1}{2}^-$ | | $\Lambda$ (1405) | | | | $(70, 1^-)$ | 1 |
| $\frac{1}{2}^+$ | $N$ (1460) | $\Lambda$ (1670) | $\Sigma$ (1560) | $\Xi$ (1825) | | $(56, 0^+)$ | 8 |
| $\frac{3}{2}^-$ | | $\Lambda$ (1520) | | | | $(70, 1^-)$ | 1 |
| $\frac{1}{2}^-$ | $N$ (1525) | $\Lambda$ (1670) | $\Sigma$ (1616) | $\Xi$ (1788) | | $(70, 1^-)$ | 8 |
| $\frac{1}{2}^-$ | $\Delta$ (1630) | | $\Sigma$ (1760) | $\Xi$ (1890) | $\Omega$ (2020) | $(70, 1^-)$ | 10 |
| $\frac{3}{2}^-$ | $\Delta$ (1670) | | | | | $(70, 1^-)$ | 10 |
| $\frac{1}{2}^-$ | $N$ (1715) | $\Lambda$ (1750) | $\Sigma$ (1690) | $\Xi$ (1755) | | $(70, 1^-)$ | 8 |
| $\frac{3}{2}^-$ | $N$ (1755) | | | $\Xi$ (1820) | | $(70, 1^-)$ | 8 |
| $\frac{5}{2}^-$ | $N$ (1675) | $\Lambda$ (1830) | $\Sigma$ (1765) | $\Xi$ (1930) | | $(70, 1^-)$ | 8 |
| $\frac{5}{2}^+$ | $N$ (1688) | $\Lambda$ (1820) | $\Sigma$ (1915) | $\Xi$ (2030) | | $(56, 2^+)$ | 8 |
| $\frac{7}{2}^+$ | $\Delta$ (1940) | | $\Sigma$ (2030) | $\Xi$ (2250) | $\Omega$ (2405) | $(56, 2^+)$ | 10 |
| $\frac{7}{2}^-$ | | $\Lambda$ (2100) | | | | $(70, 3^-)$ | 1 |
| $\frac{9}{2}^-$ | | | $\Sigma$ (2250) | $\Xi$ (2500) | | $(70, 3^-)$ | 8 |
| $\frac{9}{2}^+$ | | $\Lambda$ (2350) | $\Sigma$ (2450) | | | $(56, 4^-)$ | 8 |
| $\frac{11}{2}^+$ | $\Delta$ (2420) | | $\Sigma$ (2600) | $\Xi$ (2780) | $\Omega$ (2960) | $(56, 4^-)$ | 10 |
| $\frac{15}{2}^+$ | $\Delta$ (2850) | | $\Sigma$ (3000) | $\Xi$ (3150) | $\Omega$ (3300) | $(56, 6^-)$ | 10 |

A Tabela 8 apresenta os estados bariônicos fundamentais e alguns excitados, também em momento angular, formados pelos quarks $u$, $d$ e $s$ [5]. Para bárions formados apenas pelos quarks $u$ e $d$, os valores permitidos para spin e isospin são $\frac{1}{2}$ e $\frac{3}{2}$. Entretanto, caso o quark $s$ também seja considerado, podemos ter também bárions com isospin $I=0$ ou $I=1$. Os estados bariônicos de menor energia encontrados experimentalmente têm momento angular orbital $L=0$. Observe que nas quatro primeiras linhas da Tabela 8 temos estes bárions fundamentais já descritos nas Figuras 3 e 4, e nas Tabelas 5 e 6 da seção IV.

Recentemente, alguns estados ligados quarkiônicos exóticos, já previstos pelo Modelo de Quarks, chamados de sistemas multiquarks, foram observados em vários experimentos de altas energias. Por outro lado, há grupos que afirmam não ter visto tais estados multiquarks em seus respectivos experimentos. Na próxima seção vamos descrever, através do Modelo de Quarks, estes estados exóticos.

## VII Multiquarks: Estados Exóticos

As partículas, bárions e mésons, estudadas até o momento são pacotes de dois ou três quarks. Entretanto, nos últimos anos, houve uma procura por partículas exóticas, os chamados tetraquarks e pentaquarks, que teriam uma estrutura de dois quarks e dois antiquarks, para os tetraquarks, e quatro quarks e um antiquark, para os pentaquarks [4]. Os tetraquarks seriam basicamente estados ligados de dois mésons, enquanto os pentaquarks seriam estados ligados de um bárion com um méson. Estas ressonâncias dos estados fundamentais, previstas pelo Modelo de Quarks, preservam os princípios da QCD, pois elas têm as mesmas propriedades das partículas já estudadas, por exemplo, são sem cor.

A existência de pentaquarks foi prevista por Diakonov, Petrov e Polyakov [15] em 1997. Uma das partículas prevista por eles era o $\Theta^+$, uma partícula com massa de repouso de 1540 MeVs, com uma largura de 15 MeVs e formada pelos quarks $uudd\bar{s}$. A partícula $\Theta^+$ seria membro de um antidecupleto de bárion, representado por um triângulo como na Figuras 3 e 4, cujas extremidades seriam preenchidas por estados exóticos. Neste modelo $\Theta^+$ seria um estado ligado de um nêutron com um méson $K^+$, ou de um próton com um méson $K^0$, ou seja, $\Theta^+ = p\,K^0$ ou $\Theta^+ = n\,K^+$. Em 2003, Nakano e colaboradores [16], no Laser-Electon Photon facility SPring-8, em Osaka, Japão, também conhecido com LEPS, afirmaram ter observado $\Theta^+$. Neste experimento fótons de altíssima energia interagiriam com um nêutron, ou um próton, formando um pentaquark. Enfim, o pentaquark decairia rapidamente, em $10^{-20}$ segundos, em um méson e num nêutron ou próton. Até o momento, uma dezena de experimentos, afirmam ter constatado a existência da partícula $\Theta^+$, enquanto, outra dezena de experimentos, não observou tal estado exótico [17]. Novos experimentos estão sendo montados para elucidar esta controvérsia. A Figura 6 esquematiza como estes sistemas multiquarks seriam observados em reações de fótons, neutrinos e prótons contra alvos de nêutrons ou prótons.

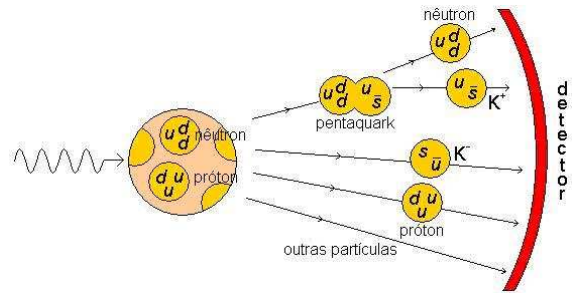

Figura 6 – Detecção das partículas usadas para reconstruir o estado de pentaquark $\Theta^+$. A formação do estado ressonante pentaquark é detectada devido ao seu decaimento em um nêutron $n$ e num méson $K^+$. Nas reações que contém káons neutros, os pentaquarks podem ser identificados através do decaimento $K^0 \to \pi^+\pi^-$.

Outros experimentos afirmam ter observado outros estados de pentaquarks, contudo a existência destas partículas ainda está em debate. Citamos o pentaquark $\Xi^{--}$, membro do antidecupleto de Diakonov, Petrov e Polyakov, formado pelos quarks $ddss\bar{u}$. O $\Xi^{--}$ seria descrito como estados ligados bárion-méson da forma: $\Xi^{--} = \Xi^-\pi^-$ ou $\Xi^{--} = \Sigma^- K^-$. A massa observada para o pentaquark $\Xi^{--}$, em colisões próton-próton no CERN, Suíça, por Alt e colaboradores [18] em 2004, foi de 1862 MeVs com uma largura de 18 MeVs. Pentaquarks compostos de mésons $D = c\bar{u}, \bar{c}u, c\bar{d}, \bar{c}d$ também foram observados. Aktas e colaboradores [19] em 2004 detectaram o pentaquark $\Theta_c = D^-\,p$ ou $\Theta_c = D^+\,\bar{p}$, em colisões elétron-próton no acelerador HERA, em Hamburgo, Alemanha. O $\Theta_c$ é formado pelos quarks $uudd\bar{c}$ e as observações indicaram uma massa de 3099 MeVs com a largura de 12 MeVs.

Em 2003, Jaffe e Wilczek [20] estenderam o modelo de pentaquarks, prevendo também os tetraquarks, que seriam estados ligados de dois mésons. Vários experimentos recentes relatam possíveis observações destes estados. Em 2003, observações de Aubert e colaboradores [21] em colisões elétron-antielétron no SLAC, Califórnia, EUA, indicaram a existência do tetraquark $D_{sJ}$. O tetraquark $D_{sJ}$ é formado por mésons $D_s = c\bar{s}, \bar{c}s$ e $\pi^0 = (u\bar{u} - d\bar{d})/\sqrt{2}$. Medidas indicaram que a massa do tetraquark $D_{sJ}$ é de 2317 MeVs com uma largura muito estreita. Enfim, outro estado tetraquark observado foi o $X$ com massa de 3872 MeVs. O tetraquark $X$ foi observado em 2003 em colisões elétron-antielétron no acelerador KEKB, em Tsukuba, Japão [22].

Estas descobertas mostram que o número de quarks que podem constituir uma partícula não são apenas três quarks (bárions) ou pares de quark-antiquark (mésons). Trabalhos recentes procuram novos modelos para descrever a força entre quarks [23]. Na referência [4], Okiharu, Suganuma e Takahashi fazem uma revisão do potencial interquarks, com ênfase no potencial de sistemas tetraquarks. Seus cálculos teóricos, utilizando uma abordagem chamada grade-QCD, onde o espaço é simulado por uma rede discreta de ponto e as equações da QCD são resolvidas por aproximações sucessivas nos pontos da rede, mostram que os tetraquarks comportam-se como um estado ligado de dois mésons.

Por outro lado, Riordan e Zajc, em [24], fazem uma revisão dos resultados experimentais obtidos nos últimos cinco anos, em experimentos de colisões ultra-relativísticas frontais de feixes de núcleos de átomos de ouro, conduzidos no acelerador RHIC, em Long Island, EUA. Verificou-se que nestas colisões a matéria comporta-se como um amontoado de quarks, glúons, fótons, elétrons e outras partículas livres, como previsto pela QCD. Na QCD a força entre quarks diminui com a distância, propriedade esta chamada de liberdade assintótica. A surpresa apresentada por estes experimentos foi que o amontoado de partículas tem também um comportamento coletivo, como um líquido, e não o de um gás quase ideal como os físicos teóricos imaginavam. Isto significa que os quarks e os glúons, mesmo no regime de liberdade assintótica da QCD, interagem coletivamente de maneira intensa, provavelmente devido à grande quantidade de partículas (congestionadas) no pequeno volume em torno da colisão. Medidas sugerem que este líquido flui quase sem viscosidade, de modo que, provavelmente, este seja o fluido mais perfeito já observado. Sobre as propriedades do plasma de quarks e glúons, em colisões realizadas no CERN/SPS e em outros experimentos no BNL/RHIC, Kodama, em [25], apresenta uma revisão dos resultados obtidos.

Em 2008 o detector ALICE [26] do acelerador LHC, no CERN, Suíça, entrará em operação. As colisões frontais de núcleos de chumbo atingirão uma energia combinada de 1 milhão de GeVs, o que permitirá uma melhor investigação das propriedades do sistema quark-glúon. Espera-se que a descoberta destes sistemas multiquarks, e a entrada em funcionamento de novos aceleradores, revelem novos aspectos da física hadrônica, especialmente aqueles relacionados com as propriedades da força interquark.

## Agradecimentos

## Referências